\documentclass[%
12pt,
aip,
cha,
amsmath,amsfonts,amssymb,
author-numerical,
reprint,
floatfix,
]{revtex4-2}

\usepackage[utf8]{inputenc}
\usepackage[T1]{fontenc}
\usepackage{mathtools}
\usepackage{graphicx}
\usepackage{hyperref}

\renewcommand{\d}[1]{\ensuremath{\operatorname{d}\!{#1}}}
\newcommand{\D}[1]{\ensuremath{\operatorname{D}\!{#1}}}

\DeclareMathOperator{\Id}{\mathsf{Id}}%
\DeclareMathOperator{\C}{\mathsf{C}}%
\DeclareMathOperator{\T}{\mathsf{T}}%

\def\A{\texttt{AVISO+}}
\def\H{Horizon Marine, Inc}
\def\E{\texttt{EddyWatch}\textsuperscript{\ttfamily\textregistered}}

\begin{document}

\title{Genesis, evolution, and apocalypse of Loop Current rings}

\author{F.\ Andrade-Canto} \email{andcanfer@gmail.com} \affiliation{
Instituto de Investigaciones Oceanol\'ogicas, Universidad
Aut\'onoma de Baja California, Ensenda, Baja California, M\'exico}

\author{D.\ Karrasch} \email{karrasch@ma.tum.de} \affiliation{Technische Universit\"at
M\"unchen, Zentrum Mathematik, Garching bei M\"unchen, Germany}

\author{F.\ J.\ Beron-Vera} \email{fberon@miami.edu} \affiliation{Department of Atmospheric
Sciences, Rosenstiel School of Marine and Atmospheric Science,
University of Miami, Miami, Florida, USA}

\date{\today}

\begin{abstract}
  We carry out assessments of the life cycle of Loop Current vortices,
  so-called rings, in the Gulf of Mexico by applying three objective
  (i.e., observer-independent) coherent Lagrangian vortex detection
  methods on velocities derived from satellite altimetry measurements
  of sea-surface height (SSH).  The methods reveal material vortices
  with boundaries that withstand stretching or diffusion, or whose
  fluid elements rotate evenly.  This involved a technology advance
  that enables framing vortex genesis and apocalypse robustly and
  with precision.  We find that the stretching- and diffusion-withstanding
  assessments produce consistent results, which show large discrepancies
  with Eulerian assessments that identify vortices with regions
  instantaneously filled with streamlines of the SSH field. The
  even-rotation assessment, which is vorticity-based, is found to
  be quite unstable, suggesting life expectancies much shorter than
  those produced by all other assessments.
\end{abstract}


\maketitle


\section{Introduction}

The Loop Current System, namely, the Loop Current itself and the
anticyclonic (counterclockwise) mesoscale (100--200-km radius)
vortices, so-called rings, shed from it, strongly influences the
circulation, thermodynamics, and biogeochemistry of the Gulf of
Mexico (GoM).\citep{Sturges-05}  As important long-range carriers,
westward-propagating Loop Current rings (LCRs) provide a potential
mechanism for the remote connectivity between the GoM's western
basin and the Caribbean Sea.\citep{Lewis-Kirwan-87, Forristall-etal-92,
Kuznetsov-etal-02, Lipphardt-etal-08, Andrade-etal-13, Donohue-etal-16,
Beron-etal-18} In particular, bringing warm Caribbean Sea water
within, the heat content of the LCRs is believed to be as significant
as for LCRs to promote the intensification of tropical cyclones
(hurricanes).\citep{Shay-etal-00} On the other hand, as regions of
strong flow shear, LCRs may be capable of producing structural
damage on offshore oil drilling rigs.\citep{Kantha-14} For all these
reasons, LCRs are routinely monitored.\citep{Sturges-Leben-00}

LCRs leave footprints in satellite altimetric sea-surface height
(SSH) maps,\citep{Leben-05} so sharp that the routine detection of
LCRs consists in the identification of regions filled with closed
streamlines of the SSH field assuming a geostrophic balance, a
practice widely followed in oceanography.\citep{Chelton-etal-11a}
However, this eddy detection approach uses instantaneous Eulerian
information to reach long-term conclusions about fluid (i.e.,
Lagrangian) transport, which are invariably surrounded by uncertainty
due to the unsteady nature of the underlying flow.  At the heart
of the issue with Eulerian eddy diagnostics of this type is their
dependence on the observer viewpoint:\citep{Peacock-etal-15} they
give different results for observers that rotate and translate
relative to one another.  The issue is most easily grasped by
bringing up, one more time as we believe is central yet widely
overlooked, the example first discussed by \citet{Haller-05} and
thereafter by others.\citep{Beron-etal-13, Haller-Beron-13,
Haller-etal-16, Haller-15}  Consider the exact solution to the
Navier--Stokes equation in two spatial dimensions: $v(x,t) =
(x^1\sin4t + x^2(2+\cos4t), x^1(\cos4t-2) - x^2\sin4t)$, where $x
= (x^1,x^2)\in \mathbb R^2$ denotes position and $t\in\mathbb R$
is time.  The flow streamlines are closed at all times suggesting
an elliptic structure (i.e., a vortex). However, this flow actually
hides a rotating saddle (pure deformation), as it follows by making
$(x^1,x^2)\mapsto (\bar x^1,\bar x^2) = (x^1\cos2t - x^2\sin2t,
x^1\sin2t + x^2\cos2t)$, under which $v(x,t) \mapsto (\bar x^2,\bar
x^1) \equiv \bar v(\bar x)$. In other words, the de-facto oceanographic
eddy detection diagnostic \citep{Chelton-etal-11a} misclassifies
the flow as vortex-like. The $\bar x$-frame is special inasmuch the
flow in this frame is steady, and thus flow streamlines and fluid
trajectories coincide.  Hence short-term exposition pictures of the
velocity field by the observer in the $\bar x$-frame determine the
long-term fate of fluid particles.  The only additional observation
to have in mind to fully determine the Lagrangian motion is that
the observer in the $\bar x$-frame rotates (at angular speed $2$).
This tells us that the flow under consideration is not actually
unsteady as there is a frame ($\bar x$) in which it is steady. In
a truly unsteady flow there is no such distinguished observer for
whom the flow is steady.\citep{Lugt-79}  Thus one can never be sure
which observer gives the right answer when the de-facto oceanographic
eddy detection diagnostic \citep{Leben-05, Chelton-etal-11a} is
applied.  As a consequence, neither false positives nor false
negatives can be ruled out,\citep{Beron-etal-13} and thus the
significance of life expectancy estimates is unclear.

Our goal here is to carry out objective (i.e, observer-independent)
assessments of the life cycle of LCRs.  This will be done in line
with recent but growing work that makes systematic use of geometric
tools from nonlinear dynamics to frame vortices objectively.
\citep{Haller-Beron-12, Beron-etal-13, Haller-Beron-13, Haller-Beron-14,
Karrasch-etal-14, Haller-etal-16, Beron-etal-19-PNAS, Haller-15,
Haller-16, Serra-Haller-16, Hadjighasem-etal-17, Haller-etal-18,
Haller-etal-19}  We will specifically apply three methods which
define \emph{coherent Lagrangian vortex} boundaries as material
loops that (i) defy stretching,\citep{Haller-Beron-13, Haller-Beron-14}
(ii) resist diffusion,\citep{Haller-etal-18} and (iii) whose elements
rotate evenly,\citep{Haller-etal-16} respectively. The fluid enclosed by such
loops can be transported for long distances without noticeable
dispersion.\citep{Wang-etal-15, Wang-etal-16}

The rest of the paper is organized as follows. In the next section
we briefly review the formal definition of each of the above coherent
Lagrangian vortex notions.  In Sec.\ \ref{sec:gen} we present a
technology that enables framing vortex genesis and apocalypse
robustly and with precision. Section \ref{sec:dat} presents the
data (satellite altimetry) on which our assessments of the life
cycle of LCRs are applied.  It also presents numerical details of
the implementation of the vortex detection methods, and introduces
the databases which deliver Eulerian assessments of the ``birth''
and ``decease'' dates of LCRs, which are used for reference.  The
results of our study are presented in Section \ref{sec:res}.  Finally,
concluding remarks are offered in Section \ref{sec:con}.

\section{Coherent Lagrangian ring detection}\label{sec:met}

Consider
\begin{equation}
  F_{t_0}^t : x_0 \mapsto x(t;x_0,t_0),
\end{equation}
the flow map resulting from integrating a two-dimensional incompressible
velocity field, namely, $v(x,t) = \nabla^\perp\psi(x,t)$ ($x\in\mathbb
R^2$ and $t\in\mathbb R$ as stated above), where $\psi$ denotes
sea-surface height. If the pressure gradient force is exclusively
due to changes in the SSH field, the latter is given by $g^{-1}f\psi(x,t)$,
where $g$ is gravity and $f$ is the Coriolis parameter, assuming a
quasigeostrophic balance.

\subsection{Null-geodesic (NG) rings}

Following \citet{Haller-Beron-13, Haller-Beron-14} we aim to identify
fluid regions enclosed by exceptional material loops that \emph{defy
the typical exponential stretching experienced by generic material
loops in turbulent flows}. This is achieved by detecting loops with
small annular neighborhoods exhibiting no leading-order variation
in averaged material stretching.

These considerations lead to a variational problem whose solutions
are loops such that any of their subsets are stretched by the same
factor $\lambda>0$ under advection by the flow from $t_0$ to $t_0
+ T$ for some $T$.  The time-$t_0$ positions of such uniformly
$\lambda$-stretching material loops turn out to be limit cycles of
one of the following two bidirectional vector or \emph{line} fields:
\begin{widetext}
\begin{equation}
  \eta_\lambda^\pm(x_0) \coloneqq
  \sqrt{
  \frac
  {\lambda_2(x_0) - \lambda^2}
  {\lambda_2(x_0) - \lambda_1(x_0)}
  }
  \,\xi_1(x_0) 
  \pm
  \sqrt{
  \frac
  {\lambda^2 - \lambda_1(x_0)}
  {\lambda_2(x_0) - \lambda_1(x_0)}
  }
  \,\xi_2(x_0),  
  \label{eq:eta}
\end{equation}
\end{widetext}
where $\lambda_1(x_0) < \lambda^2 < \lambda_2(x_0)$.  Here,
$\{\lambda_i(x_0)\}$ and $\{\xi_i(x_0)\}$ satisfying
\begin{equation}
  0 < \lambda_1(x_0) \equiv
  \frac{1}{\lambda_2(x_0)} < 1,\quad 
  \langle\xi_i(x_0),\xi_j(x_0)\rangle = \delta_{ij},
   \label{eq:eig}
\end{equation}
$i,j = 1,2$, are eigenvalues and (orientationless) normalized
eigenvectors, respectively, of the Cauchy--Green (strain) tensor,
\begin{equation}
  \C_{t_0}^{t_0+T}(x_0) \coloneqq \D{F}_{t_0}^{t_0+T}(x_0)^\top
  \D{F}_{t_0}^{t_0+T}(x_0).  
  \label{eq:C}
\end{equation}
The tensor field $\C_{t_0}^{t_0+T}(x_0)$ objectively measures
material deformation over the time interval $[t_0,t_0+T]$. Limit
cycles of \eqref{eq:eta} or \emph{$\lambda$-loops} either grow or
shrink under changes in $\lambda$, forming smooth annular regions
of non-intersecting loops.  The outermost member of such a band of
material loops is observed physically as the boundary of a
\emph{coherent Lagrangian ring}. The $\lambda$-loops can also be
interpreted as so-called null-geodesics of the indefinite tensor
field $\C_{t_0}^{t_0+T}(x_0)-\lambda\Id$, which is why we also refer
to them as \emph{null-geodesic} (or \emph{NG}) \emph{rings}.

\subsection{Diffusion-barrier (DB) rings}

Another recent approach to coherent vortices in geophysical flows
has been put forward in \citet{Haller-etal-18}. In this case one
aims at identifying fluid regions that \emph{defy diffusive transport
across their boundaries}. Note that by flow invariance, any fluid
region has vanishing advective transport across its boundary. In
turbulent flows, however, a generic fluid region has massive
\emph{diffusive} leakage through its boundary, which correlates
with the typical exponential stretching of the latter.

A technical challenge is that the diffusive flux of a virtual
diffusive tracer through a material surface over a finite time
interval $[t_0,t_0+T]$ depends on the concrete evolution of the
scalar under the advection--diffusion equation.  In the limit of
vanishing diffusion, however, \citet{Haller-etal-18} show that the
diffusive flux through a material surface can be determined by the
gradient of the tracer at the initial time instance and a tensor
field $\T$ that can be interpreted as the time average of the
diffusion tensor field along a fluid trajectory. In the case of
isotropic diffusion, this reduces to the average of inverse
Cauchy--Green tensors,
\begin{equation}\label{eq:transporttensor}
  \T(x_0) \coloneqq \frac{1}{T}\int_{t_0}^{t_0+T}
  \left(\C_{t_0}^{t_0+t}(x_0)\right)^{-1}\d{t}.
\end{equation}

Searching for material loops with small annular neighborhoods
exhibiting no leading-order variation in the vanishing-diffusivity
approximation of diffusive transport leads to a variational problem
whose solutions are limit cycles of \eqref{eq:eta}, where now
$\lambda_i$ and $\xi_i$ are, respectively, eigenvalues and eigenvectors
of the time-averaged Cauchy--Green tensor
\begin{equation}\label{eq:avCG}
    \bar{\C}_{t_0}^{t_0+T}(x_0) \coloneqq
    \frac{1}{T}\int_{t_0}^{t_0+T} \C_{t_0}^{t_0+t}(x_0)\d{t}.
\end{equation}
This simple tensor structure assumes isotropic diffusion and an
incompressible fluid flow. We refer to vortices obtained by this
methodology as \emph{diffusion-barrier} (or \emph{DB}) \emph{rings}.
Due to the mathematical similarity to the geodesic ring approach,
we may use the same computational method as for NG rings, simply
by replacing $\C_{t_0}^{t_0+T}$ by $\bar{\C}_{t_0}^{t_0+T}$.

\subsection{Rotationally-coherent (RC) rings}

In our analysis, we also employ a third methodology, which was
developed by \citet{Haller-etal-16}. It puts less emphasis on
specific properties of the boundary (like stretching or diffusive
flux) of coherent vortices, but highlights that coherent vortices
are often associated with \emph{concentrated regions of high
vorticity}. Defining vortices in terms of vorticity has a long
tradition,\citep{Okubo-70, Weiss-91} but in unsteady fluid flows
it comes with a number of drawbacks, one of which is the lack of
objectivity.\citep{Haller-15} In \citet{Haller-etal-16}, the authors
overcome these challenges by showing that the \emph{Lagrangian
averaged vorticity deviation} (or \emph{LAVD}) field
\begin{equation}
  \mathrm{LAVD}_{t_0}^{t_0+T}(x_0) \coloneqq \int_{t_0}^{t_0+T}\left\lvert
  \omega\big(F_{t_0}^t(x_0),t\big)-\bar{\omega}(t)\right\rvert\d{t},
\end{equation}
is an objective scalar field. Here, $\omega(x,t)$ is the vorticity
of the fluid velocity at position $x$ and time $t$, and $\bar{\omega}(t)$
is the vorticity at time $t$ averaged over the tracked fluid bulk.
In this framework, vortex centers are identified as maxima of the
LAVD field, and vortex boundaries as outermost convex LAVD-level
curves surrounding LAVD maxima.  Because loops are composed of fluid
elements that complete the same total material rotation relative
to the mean material rotation of the whole fluid mass, we will refer
to the vortices as \emph{rotationally-coherent} (or \emph{RC})
\emph{rings}.  In practice, the convexity requirement is relaxed,
using a ``tolerable'' convexity deficiency.\citep{Haller-etal-16}
In contrast to the two previously described methods, the LAVD
approach therefore does not address vortex boundaries directly (say,
via a variational approach), but deduces them as level-set features
of the objective LAVD field.

\section{Genesis and apocalypse}\label{sec:gen}

Our main goal is to study genesis, evolution, and apocalypse of
LCRs from an objective, Lagrangian point of view.  Since there is
no generally agreed definition of the concept of a coherent vortex,
we need to employ several proposed methods to rule out the possibility
that the results are biased by the specific choice of method.

To determine the ``birth'' or the ``decease'' of a coherent Lagrangian
vortex in a robust fashion, we need to eliminate a couple of
potentially biasing issues. First, as stated above, we include
several Lagrangian methodologies in our study. Second, we want to
avoid potential sensitivities due to implementation details (such
as algorithm or parameter choices).  Recall that Lagrangian approaches
choose not only an initial time instance $t_0$, but also a flow
horizon $T$. A naive approach to the determination of the decease
of a coherent vortex would be to simply take the maximum of $t_0+T$
for which a Lagrangian method detects a coherent vortex, where $t_0$
and $T$ are taken from a range of reasonable values. While this
approach yields a definite answer, it may be totally inconsistent
with other computations run for different choices of $t_0$ and $T$.
For instance, if a Lagrangian computation detects a coherent vortex
over the time interval $[t_0, t_0+T]$, it should also detect a
vortex over the time interval $[t_0+\delta t, (t_0+\delta t) +
(T-\delta t)] = [t_0+\delta t, t_0+T]$ for small $|\delta t|$, if
$t_0+T$ was really the date of breakdown.

In order to make our predictions statistically more robust and prove
internal consistency, we employ the following approach. First, we
run Lagrangian simulations on a temporal double grid as follows.
We roll the initial time instance $t_0$ over a time window roughly
covering the time interval of vortex existence, which we seek to
determine. For each $t_0$, we progress $T$ in 30-day steps as long
as the Lagrangian method successfully detects a coherent vortex.
Thus, we obtain for each $t_0$ a \emph{life expectancy} $T_{\max}(t_0)$,
which is the maximum $T$ for which a Lagrangian simulation starting
at $t_0$ successfully detected a coherent vortex.

Ideally, we would like to see the following $T_{\max}(t_0)$ pattern.
Assume a coherent Lagrangian vortex breaks down on day 200, counted
from day 0. Then for $t_0 = 0$ the longest successful vortex detection
should yield a $T_{\max}(0) = 180$ d.  Similarly, for $t_0 = 5, 10,
15, 20$ we should get a $T_{\max}=180$ d.  From $t_0=25$ on, however,
we should start seeing $T_{\max}$ dropping down to $150$ days,
because for $T=180$ days, the Lagrangian flow horizon reaches beyond
the vortex breakdown.  As a consequence, we would like to see a
wedge-shaped $T_{\max}(t_0)$ distribution, which would indicate
that all Lagrangian coherence assessments predict the breakdown
consistently, though slightly smeared out regarding the exact date.
If encountered, such a consistent prediction of breakdown would
arguably remove the possibility of degenerate results. To summarize,
in an ideal case, a Lagrangian simulation of the lifespan of a
coherent vortex would therefore start with a large $T_{\max}$-value,
which consistently decreases as $t_0$ progresses forward in time.

It turns out that in many cases such wedge-shaped $T_{\max}(t_0)$-patterns
can be indeed observed, sometimes with astonishing clarity,
given the finitely resolved velocity fields and the complexity of
the Lagrangian calculation and vortex detection algorithms.

\section{Data and numerical implementation}\label{sec:dat}

The SSH field from which the flow is derived is given daily on a
0.25$^{\circ}$-resolution longitude--latitude grid.  This represents
an absolute dynamic topography, i.e., the sum of a (steady) mean
dynamic topography and the (transient) altimetric SSH anomaly.  The
mean dynamic topography is constructed from satellite altimetry
data, in-situ measurements, and a geoid model.\citep{Rio-Hernandez-04}
The SSH anomaly is referenced to a 20-yr (1993--2012) mean, obtained
from the combined processing of data collected by altimeters on the
constellation of available satellites.\citep{LeTraon-etal-98}

Computationally, we detect NG and DB rings from the altimetry-derived
flow by the method devised in \citet{Karrasch-etal-14} and recently
extended for large-scale computations in Karrasch and
Schilling,\citep{Karrasch-Schilling-20} as implemented in the package
\texttt{CoherentStructures.jl}.  It is written in the modern
programming language \texttt{Julia}, and is freely available from
\texttt{https://\allowbreak github.com/\allowbreak
CoherentStructures/\allowbreak CoherentStructures.jl}. In turn, RC
ring detection, computationally much more straightforward, was
implemented in \texttt{MATLAB}\textsuperscript{\ttfamily\textregistered}
as described in \citet{Beron-etal-19-PNAS} (a software tool, not
employed here, is freely distributed from \texttt{https://\allowbreak
github.com/\allowbreak LCSETH/\allowbreak Lagrangian-\allowbreak
Averaged-\allowbreak Vorticity-\allowbreak Deviation-\allowbreak
LAVD}).  The spacing of the grid of initial trajectory positions
in all cases is set to 0.1 km as in earlier Lagrangian coherence
analyses involving altimetry data.  \citep{Beron-etal-13,
Olascoaga-etal-13, Beron-etal-18} Trajectory integration is
carried out using adaptive time-stepping schemes and involves
cubic interpolation of the velocity field data.  NG and DB rings
are sought with stretching parameter ($\lambda$) ranging over the interval
$\lambda \in [1 \pm 0.5]$.  Recall that $\lambda = 1$ NG-vortices
reassume their arc length at $t_0+T$.\citep{Haller-Beron-13} When
the flow is incompressible (as is the case of the altimetry-derived
flow) such $\lambda = 1$ vortices stand out as the most coherent
of all as their boundaries resist stretching while preserving the
area they enclose.  Following \citet{Haller-etal-16} the convexity
deficiency is set to $10^{-3}$ for the RC ring extractions.

As our interest is in LCRs, we concentrate on the time intervals
on which these were identified by \H.\ as part of the \E\, program.
This program identifies LCRs as regions instantaneously filled with
altimetric SSH streamlines.\citep{Leben-05}  The \E\, program has
been naming LCRs and reporting their birth and decease dates since
1984.  Our analysis is restricted to the period 2001--2013, long
enough to robustly test theoretical expectations and for the results
to be useful in applications such as ocean circulation model
validation.

Alternative assessments of the genesis and apocalypse of LCRs are
obtained from the \A\, Mesoscale Eddy Trajectory Atlas Product,
which is also computed from the Eulerian footprints left by the
eddies on the global altimetric SSH field.\citep{Chelton-etal-11a}

\section{Results}\label{sec:res}

We begin by testing our expectation that Lagrangian life expectancy
($T_{\max}$) should decrease with increasing screening time ($t_0$),
exhibiting a wedge shape.  We do this by focusing on LCR \emph{Kraken},
so named by \E\, and recently subjected to a Lagrangian coherence
study.\citep{Beron-etal-18} In that study the authors characterized
\emph{Kraken} as an NG ring using altimetry data.  Furthermore,
they presented support for their characterization by analyzing
independent data, namely, satellite-derived color (Chl concentration)
and trajectories from satellite-tracked drifting buoys. This rules
out the possibility that LCR \emph{Kraken} is an artifact of the
satellite altimetric dataset, thereby constituting a solid benchmark
for testing our expectation.  The authors of the aforementioned
study estimated a Lagrangian lifetime for \emph{Kraken} of about
200 d, but framing the genesis and apocalypse of the ring with
precision was beyond the scope of their work.

\begin{figure*}[t!]
  \centering%
  \includegraphics[width=\textwidth]{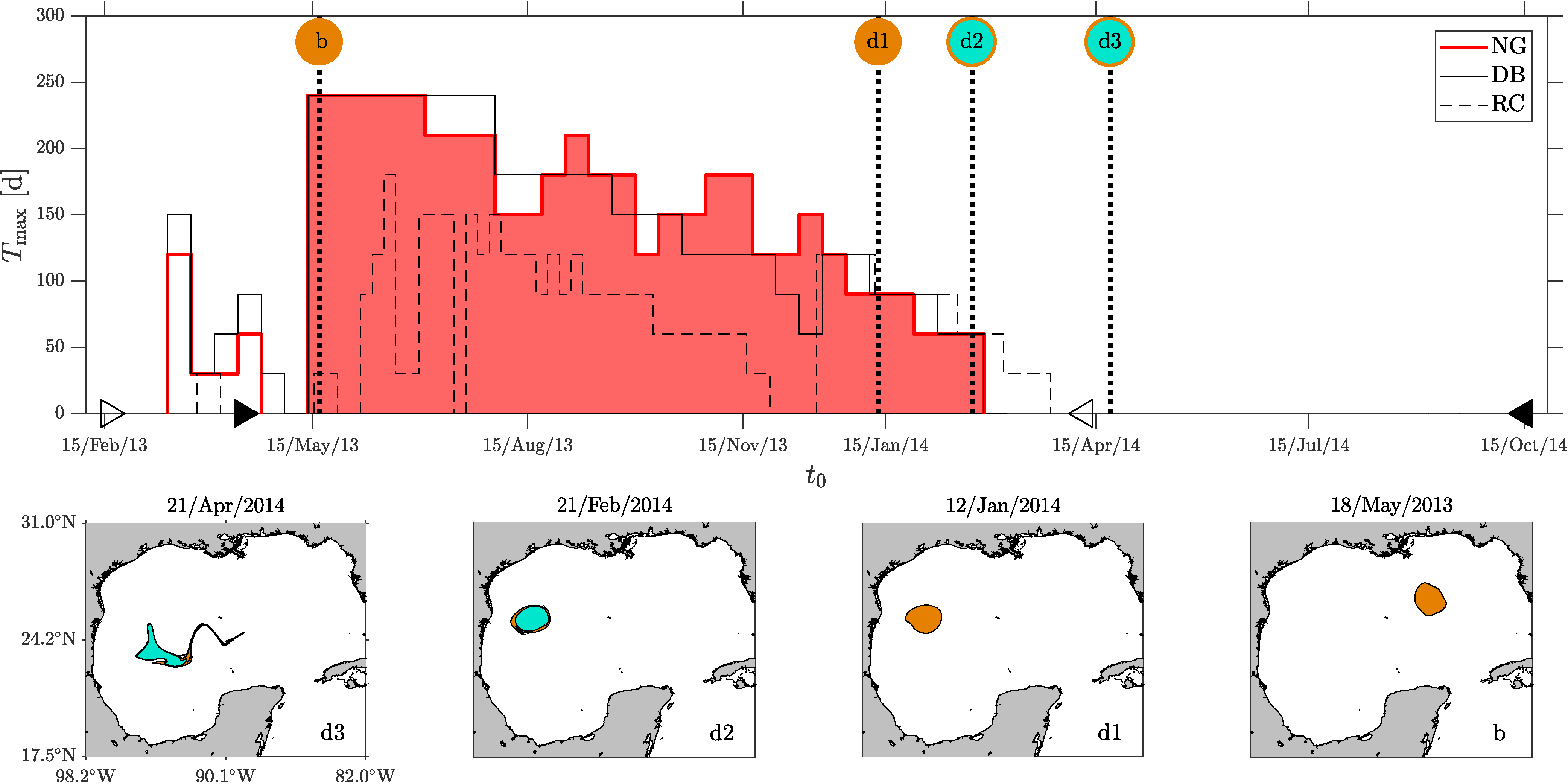} \caption{(top
  panel) For Gulf of Mexico's Loop Current ring (LCR) \emph{Kraken},
  life expectancy as a function of screening time according
  null-geodesic (NG), diffusion-barrier (DB), and rotationally-coherent
  (RC) Lagrangian vortex assessments.  Indicated are the birth date
  of the ring (b), and three decease date estimates (d1, d2, and
  d3); cf.\ text for details.  Birth and decease dates according
  to \E\, and \A\, Eulerian vortex assessments are indicated with
  open and filled triangles, respectively. (bottom panel) Based on
  the NG assessment, LCR \emph{Kraken} on birth date and the three
  decease date estimates.}
  \label{fig:kra-fwd}%
\end{figure*}

The top panel of Fig.\ \ref{fig:kra-fwd} shows $T_{\max}(t_0)$ for
\emph{Kraken} based on NG (red), DB (solid black), and RC (dashed
black) coherence assessments.  First note that the NG and DB
assessments are largely consistent, producing a wide-base $T_{\max}(t_0)$
wedge with height decreasing with increasing $t_0$ in addition to
a short, less well-defined wedge prior to it.  This becomes very
evident when compared to the RC assessment, which produces intermittent
wedge-like $T_{\max}(t_0)$ on various short $t_0$-intervals. We
have observed that this intermittency is typical, rather than
exceptional, for the RC assessment.  Thus we consider NG coherent
Lagrangian vortex detection, which in general produces nearly
identical results as DB vortex detection, in the genesis and
apocalypse assessments that follow.

Indicated in the top panel of Fig.\ \ref{fig:kra-fwd} (with a
vertical dashed line) is our estimate of the birth date of LCR
\emph{Kraken}, $t_0=$ 18/May/2013, and three estimates of its decease
date, to wit, $t_0=$ 12/Jan/2014, 21/Feb/2014, and 21/Apr/2014. The
birth date corresponds to the $t_0$ marking the leftmost end of the
$T_{\max}(t_0)$ wedge with the longest base (highlighted).  Our
first decease date estimate (d1) is given by the birth date plus its
life expectancy, set by the height of the wedge or 12/Jan/2014 $-$
18/May/2013 $=239$ d.  The second decease date estimate (d2) is given
by the $t_0$ marking the rightmost end of the wedge, which is 40-d
longer than its life expectancy.  Our third decease date estimate (d3)
is given by the second decease date estimate plus the height of the
wedge at its rightmost end, namely, 21/Apr/2014 $-$ 21/Feb/2014
$=59$ d.

The bottom panel of the Fig.\ \ref{fig:kra-fwd} shows, in orange,
LCR \emph{Kraken} on its estimated birth date (b), and its advected
image under the altimetry-derived flow on the first (d1), second
(d2), and third (d3) decease date estimates.  Overlaid on the later
on the second and third decease date estimated are (shown in cyan)
the NG vortex extracted on the second decease date estimate and its
advected image on the third decease date estimate.  Note that on
the first and second decease date estimates \emph{Kraken} does not
show any noticeable signs of outward filamentation. On the third
decease date estimate most of the original fluid mass enclosed by
the ring boundary exhibits a coherent aspect. Evidently, the first
and second decease date estimates are too conservative, so it is
reasonable to take the third one as the most meaningful decease
date estimate of the three.  We will refer to it as \emph{the}
decease date.

Indicated by open and filled triangles in the abscissa of the
$T_{\max}(t_0)$ plot in the top panel of the Fig.\ \ref{fig:kra-fwd}
are the Eulerian assessment of birth and decease dates of \emph{Kraken}
by \E, and \A, respectively.  \E\, overestimates the decease date
by about 180 d, while \A, underestimates it somewhat, by 19 d.  To
evaluate the performance of Eulerian vortex detection in assessing
the birth date of \emph{Kraken} an additional analysis is needed.

\begin{figure*}[t!]
  \centering%
  \includegraphics[width=\textwidth]{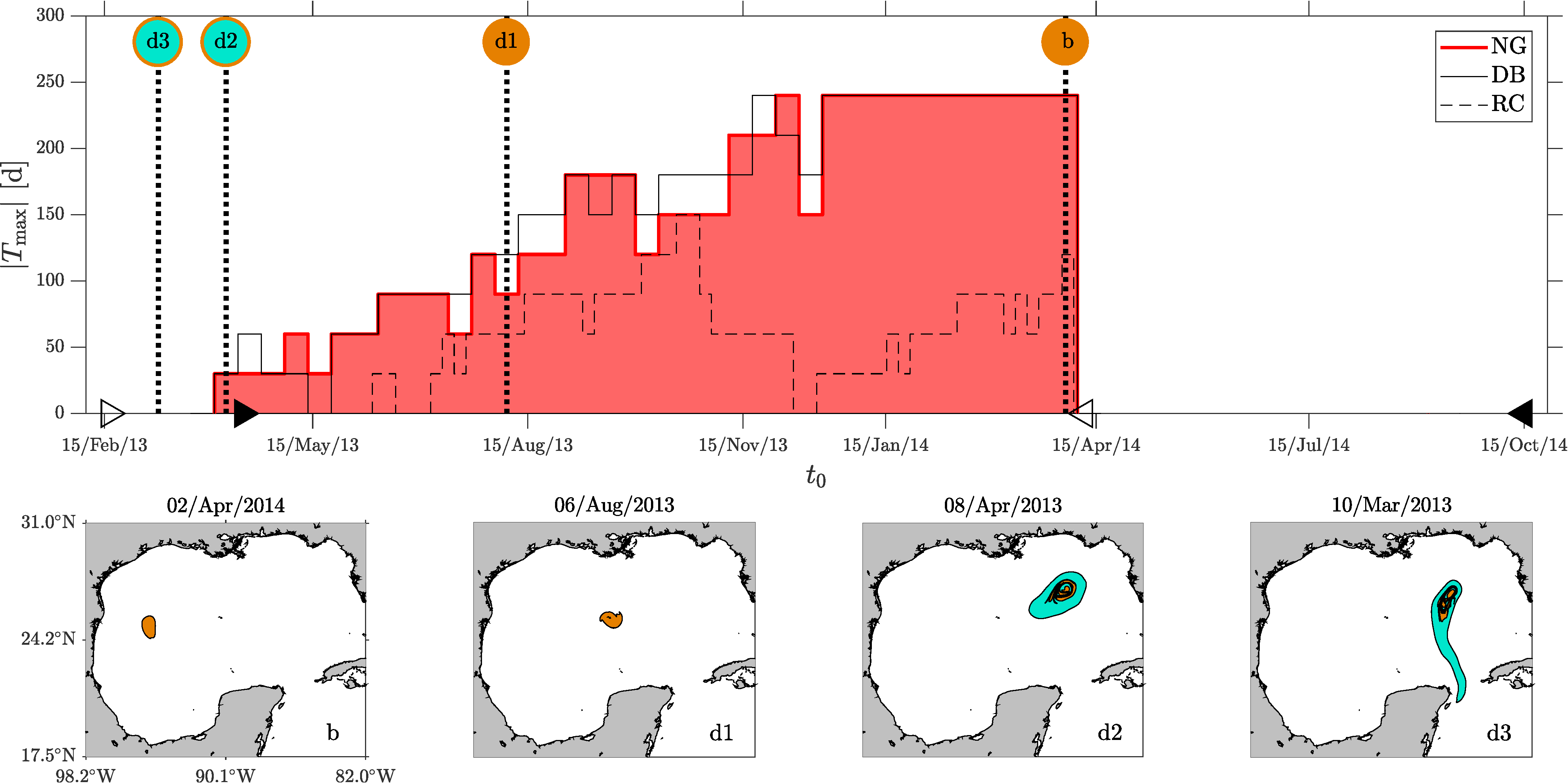} \caption{As in
  Fig.\ \ref{fig:kra-fwd}, but for assessments made in backward
  time.}
  \label{fig:kra-bwd}%
\end{figure*}

The results from such an analysis are presented in Fig.\
\ref{fig:kra-bwd}, which shows the same as in Fig.\ \ref{fig:kra-bwd}
but as obtained from applying all the Lagrangian vortex detection
methods \emph{backward} in time, i.e., with $T<0$, around the
\emph{Kraken}'s decease date.  The top panel of the figure shows
(now) $|T_{\max}|$ as a function of screening time $t_0$.  Note
that the NG and DB coherence assessments produce single wide-base
$|T_{\max}|(t_0)$ wedges with height decreasing with decreasing
$t_0$, nearly indistinguishable from one another. As in the
forward-time analysis, the RC assessment shows intermittent wedge-like
$|T_{\max}|(t_0)$ on various short $t_0$-intervals, suggesting a
much shorter life expectancy than observed in reality. Thus we turn
our attention to the NG (or DB) assessment.  This produces a
backward-time birth date estimate on $t_0=$ 02/Apr/2014, and three
backward-time decease date estimates on 06/Apr/203, 08/Apr/2013,
and 10/Mar/2013.  In forward time, 02/Apr/2014 represents a decease
date estimate, which is only 19-d earlier than the decease date
obtained above from forward-time computation.  The largest discrepancy
between forward- and backward-time assessments are seen for the
birth date.  Following the forward-time computation reasoning above,
the backward-time computations sets it 296 d earlier, on $t_0=$
10/Mar/2013. This lies 21 and about 30 d later and earlier than to
the \A\, and \E, assessments, respectively, which are instantaneous,
i.e., they do not depend on the time direction on which they are
made.

The backward-time estimate of \emph{Kraken}'s decease date can be
taken to represent a forward-time \emph{conception} date estimate
for the ring.  This is quite evident from the inspection of the
bottom panel of Fig.\ \ref{fig:kra-bwd}, which shows (in orange)
LCR \emph{Kraken} as extracted from backward-time computation on
the backward-time birth date estimate, and images thereof under the
backward-time flow on the three backward-time decease date estimates.
On the last two decease date estimates, these are shown overlaid
on the ring extracted from backward-time computation on the second
backward-time decease date estimate and its backward-advected image
on the third backward-time decease date estimate,  which represents,
as noted above, a conception date for \emph{Kraken}.

\begin{figure}[t!]
  \centering%
  \includegraphics[width=\columnwidth]{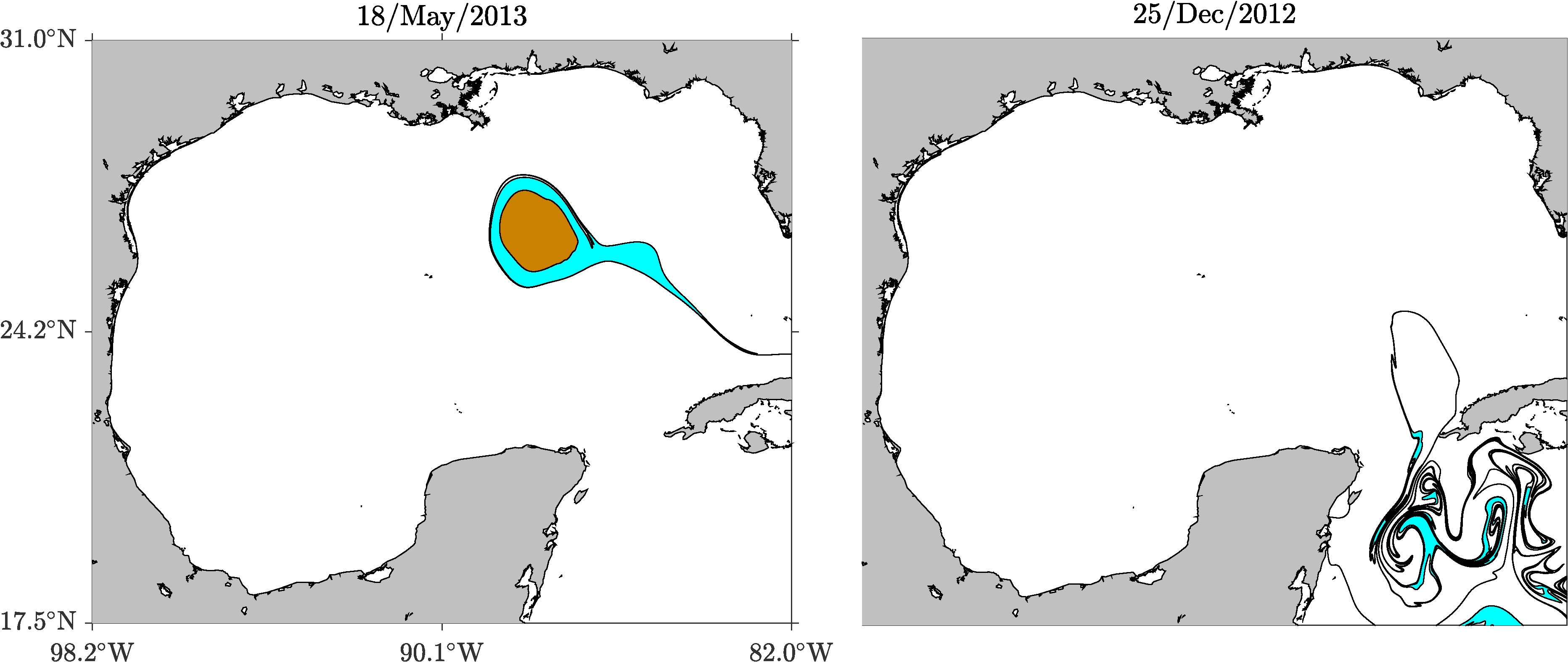}%
  \caption{(left panel) LCR \emph{Kraken} on  birth date overlaid
  in orange on the forward-advected image of the ring extracted
  from backward-time computation on 08/Apr/2013 (third backward-time
  decease estimate).  (right panel) Backward-advected image of the
  fluid region indicated in cyan in the left panel.}
  \label{fig:kra-gen}%
\end{figure}

Indeed,  the fluid region indicated in cyan contains at all times
the fluid region indicated in orange.  Thus the orange fluid is
composed of the same fluid as the cyan fluid.  Furthermore, the
cyan fluid, which can be traced back into the Caribbean Sea, ends
up forming the fluid that forms \emph{Kraken} on its (forward-time)
birth date.  This is illustrated in left panel of Fig.\ \ref{fig:kra-gen},
which shows \emph{Kraken} (in orange) as obtained from forward-time
computation on its birth date overlaid on the forward-advected image
of the cyan fluid. In the right panel we show a backward-advected
image of the cyan fluid that reveals its origin in the Caribbean
Sea. The supplementary material includes an animation (Mov.\ 1) 
illustrating the full life cycle of LCR \emph{Kraken}.

\begin{figure*}[t!]
  \centering%
  \includegraphics[width=\textwidth]{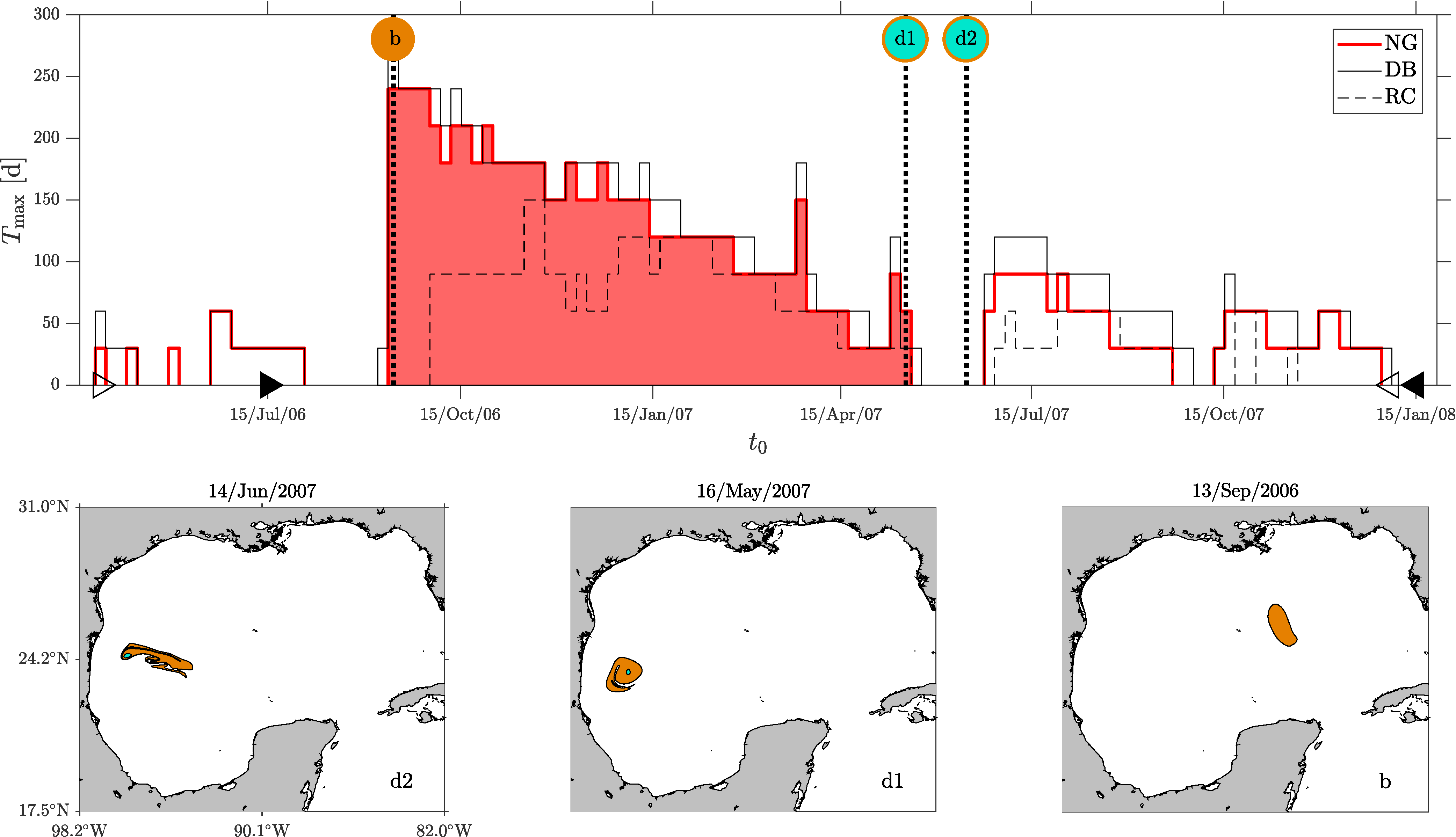}%
  \caption{As in Fig.\ \ref{fig:kra-fwd}, but for LCR \emph{Yankee}.}
  \label{fig:yan}%
\end{figure*}

We note that the need of introducing the conception date estimate
could have been anticipated from the inspection of the forward-time
assessment.  Note the short wedge-like $T(t_0)$ before the long-base
wedge in Fig.\ \ref{fig:kra-fwd} employed in assessing genesis and
apocalypse.  In a way the presence of that short wedge-like $T(t_0)$
was already insinuating that coherence was building sometime before
the ring was declared born.  Similar disconnected wedge-like $T(t_0)$
patterns may be observed past the main wedge, as can be seen in
Fig.\ \ref{fig:yan}, which shows the same as Fig.\ \ref{fig:kra-fwd}
but for LCR \emph{Yankee}. These wedge-like patterns, however, are
not signs of the ring's ``resurrection,'' but actually correspond
to vortex structures in general unrelated or only partly related
to the ring in question.

\begin{figure*}[t!]
  \centering%
  \includegraphics[width=\textwidth]{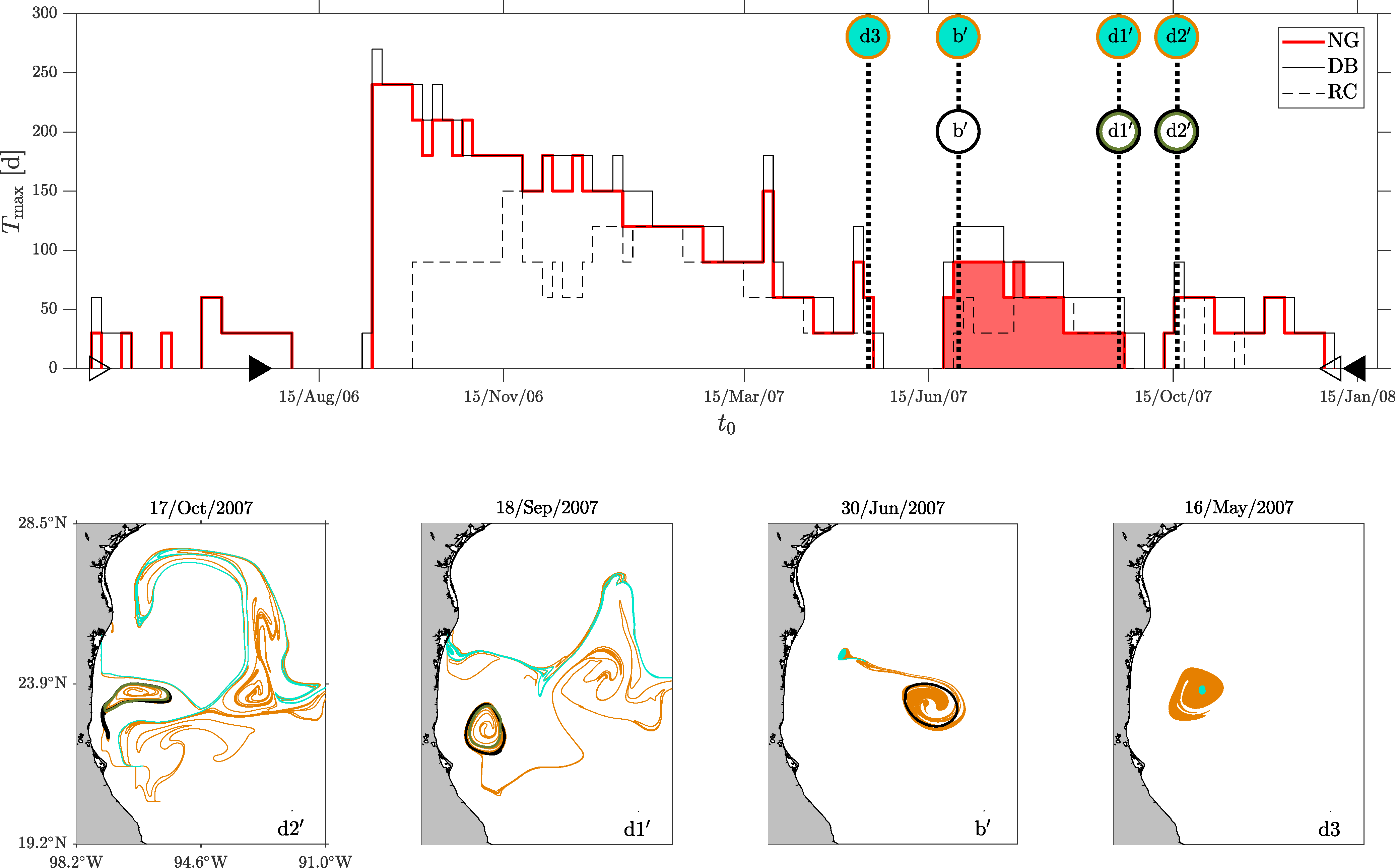} \caption{As in
  Fig.\ \ref{fig:yan}, with a focus on the highlighted piece of the
  $T_{\max}(t_0)$ plot.}
  \label{fig:cri}%
\end{figure*}

We illustrate the above in Fig.\ \ref{fig:cri}.  Note the appearance
of two short wedge-like patterns past the main $T(t_0)$ wedge.  Let
us concentrate attention on the earliest of the two short wedges.
We infer a forward-time birth date is 30/Jun/2007, and two forward-time
decease dates on 18/Sep/2017 and 17/Oct/2017.  The bottom panels
of the figure show how these characterize the life cycle of a vortex,
newly formed and composed only in part of LCR \emph{Yankee}'s fluid.
This is evident by comparing the position of the vortex on birth
date and first decease date estimates and their forward-advected
images with those of \emph{Yankee} as revealed on 13/Sep/2006.  The
\E\, and \A\, nonobjective Eulerian vortex assessments fail to frame
this, largely overestimating the decease date of \emph{Yankee}.

We compile in Table \ref{tab:cen} the objective Lagrangian estimates
of conception, birth, and decease dates for all LCRs during the
altimetry era. An entry of table left blank means that the ring
could not be classified as coherent.  The objective estimates are
compared with nonobjective Eulerian estimates by \E\, and \A, with
the former only providing the month of the year when birth and
decease take place.  Note the tendency of the Eulerian assessments
to overestimate the birth and decease dates of the rings.  Indeed,
the Eulerian assessments cannot distinguish between conception and
birth. They typically keep track of vortex-like structures past the
decease date of the rings, which, present around that date, are not
formed by the fluid mass contained by the rings.  Moreover, the
Eulerian assessments classify as coherent, and even name, rings
that turn out not to be so. The supplementary material includes
two animations supporting these conclusions for features classified
as LCRs \emph{Quick} (Mov.\ 2) and \emph{Sargassum} (Mov.\ 3) by
the \E\, and \A\, nonobjective Eulerian assessments.

\renewcommand{\arraystretch}{1.25}
\begin{table*}[t!]
  \linespread{1}\selectfont{}
  \centering
  \begin{tabular}{ccccccccccc}%
    \hline\hline%
	 &&  && \multicolumn{3}{c}{Birth date} && \multicolumn{3}{c}{Decease date}\\
	 \cline{5-7}\cline{9-11}
	 \multicolumn{1}{l}{Ring}&&\multicolumn{1}{c}{Conception date}&&Objective&\E&\A&&Objective&\E&\A\\
	 \hline
	 \multicolumn{1}{l}{\emph{Nansen}}&&27/02/01&&13/03/01&04/01&18/03/00&&31/07/01&12/01&04/01/02\\ 
	 \multicolumn{1}{l}{\emph{Odesa}}&&02/07/01&&31/07/01&09/01&23/03/01&&09/10/01&12/01&05/11/01\\ 
	 \multicolumn{1}{l}{\emph{Pelagic}}&&&&&12/01&07/09/01&&&05/02&17/02/02\\ 
	 \multicolumn{1}{l}{\emph{Quick}}&&&&&03/02&19/02/02&&&04/03&05/05/03\\ 
	 \multicolumn{1}{l}{\emph{Sargasum}}&&&&&05/03&30/03/02&&&12/03&16/01/04\\
	 \multicolumn{1}{l}{\emph{Titanic}}&&10/11/03&&09/12/03&10/03&01/08/03&&21/06/04&10/04&13/11/04\\ 
	 \multicolumn{1}{l}{\emph{Ulises}}&&22/11/04&&11/12/04&05/04&07/12/03&&15/05/05&09/05&07/10/05\\ 
	 \multicolumn{1}{l}{\emph{Extreeme}}&&&&&03/06&13/01/06&&&09/06&13/11/06\\ 
	 \multicolumn{1}{l}{\emph{Yankee}}&&04/09/06&&13/09/06&07/06&26/04/06&&31/05/07&01/08&03/01/08\\ 
	 \multicolumn{1}{l}{\emph{Zorro}}&&10/03/07&&24/03/07&04/07&26/08/06&&16/08/07&08/07&17/08/07\\ 
	 \multicolumn{1}{l}{\emph{Albert}}&&02/11/07&&01/12/07&11/07&21/03/07&&05/03/08&05/08&23/04/08\\ 
	 \multicolumn{1}{l}{\emph{Cameron}}&&20/06/08&&20/06/08&07/08&15/06/08&&10/02/09&05/09&18/06/09\\ 
	 \multicolumn{1}{l}{\emph{Darwin}}&&29/01/09&&02/02/09&12/08&13/02/08&&25/10/09&11/09&10/11/09\\ 
	 \multicolumn{1}{l}{\emph{Ekman}}&&11/04/09&&09/07/09&07/09&04/02/09&&20/05/10&03/11&28/08/10\\ 
	 \multicolumn{1}{l}{\emph{Hadal}}&&22/06/11&&11/07/11&08/11&23/11/10&&23/12/11&03/12&26/12/11\\ 
	 \multicolumn{1}{l}{\emph{Icarus}}&&08/10/11&&22/10/11&11/11&19/07/11&&11/09/12&02/13&04/10/12\\ 
	 \multicolumn{1}{l}{\emph{Jumbo}}&&28/04/12&&28/04/12&06/12&18/04/12&&21/08/12&02/13&10/11/12\\
	 \multicolumn{1}{l}{\emph{Kraken}}&&10/03/13&&08/04/13&04/13&17/02/13&&02/04/14&10/14&10/04/14\\ 
	 \hline%
  \end{tabular}%
  \caption{Objective Lagrangian estimates of conception, birth, and
  decease dates of Loop Current rings in the Gulf of Mexico identified
  from satellite altimetry over 2001--2013 along with nonobjective
  Eulerian estimates of birth and decease dates.} \label{tab:cen}
\end{table*}\renewcommand{\arraystretch}{1}

\begin{figure}[t!]
  \centering%
  \includegraphics[width=\columnwidth]{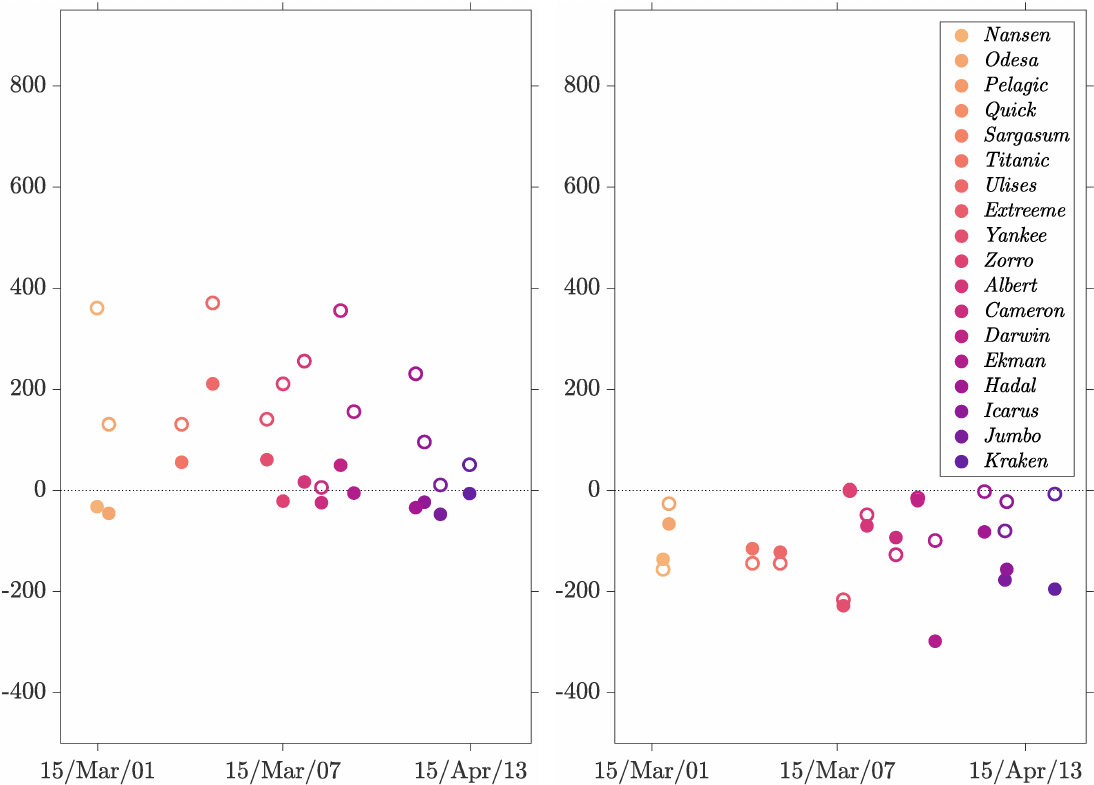}%
  \caption{As a function of time, difference (in d) between objective
  Lagrangian and nonobjective Eulerian estimates of birth (left
  panel) and death (right panel) dates. Dots (resp., circles) involve
  \E, (resp., \A) assessments.} \label{fig:diff}
\end{figure}

We conclude by highlighting the disparities between the objective
Lagrangian and nonobjective Eulerian assessments of the genesis and
apocalypse of LCR rings in Fig.\ \ref{fig:diff}.  The figure presents,
as a function of time over 2001--2011, the difference (in d) between
NG and E\, (dots), and NG and A\, (circles) assessments of birth
(left) and decease (right) dates.  The differences can
be quite large (up to 1 yr!) with Eulerian assessments, which
in general underestimate the birth dates of the rings and overestimate
their decease dates.

\section{Conclusions}\label{sec:con}

We have carried out an objective (i.e., observer-independent)
Lagrangian assessment of the life cycle of the Loop Current rings
(LCRs) in Gulf of Mexico detected from satellite altimetry.  Three
objective methods of coherent Lagrangian vortex detection were
considered here.  These reveal material vortices with boundaries
that defy stretching or diffusion, and whose elements rotate evenly.
A modest technology advance was performed which enabled framing
vortex genesis and apocalypse with robustness and precision. We
found that the stretching- and diffusion-defying assessments produce
consistent results.  These in general showed large discrepancies
with Eulerian assessments which identify vortices with regions
instantaneously filled with streamlines of the SSH field. The
Eulerian assessments were found incapable to distinguish conception
from birth of the rings.  They also tended to track past their
decease dates vortex-like features unrelated to the rings in question.
The even-rotation assessment, which is vorticity-based, was found
to be quite unstable, suggesting life expectancies much shorter
than those produced by all other assessments. The inconsistency
found adds to the list of known issues of LAVD-based vortex statistics,
\cite{Tarshish-etal-18} including high sensitivity with respect to
the choice of computational parameter values. Our results can find
value in drawing unambiguous evaluations of material transport and
should represent a solid metric for ocean circulation model
benchmarking.

\section*{Supplementary material}

The supplementary material contains three animations. Movie 1
illustrates the complete life cycle of LCR \emph{Kraken} as assessed
objectively using NG-ring detection.  Movies 2 and 3 show sequences
of advected images of the features classified as LCRs \emph{Quick}
and \emph{Sargassum}, respectively, by the \E\, and \A\, nonobjective
Eulerian assessments.

\section*{Author's contributions}

All authors contributed equally to this work.

\begin{acknowledgments}
  This work was initiated during the ``Escuela interdisciplinaria
  de transporte en fluidos geofísicos: de los remolinos oce\'anicos
  a los agujeros negros,'' Facultad de Ciencias Exactas y Naturales,
  Universidad de Buenos Aires, 5--16/Dec/2016.  Support from Centro
  Latinoamericano de Formaci\'on Interdisciplinaria is sincerely
  appreciated. This work was supported by CONACyT--SENER (Mexico)
  under Grant No.\ 201441 (FAC, FJBV) as part of the Consorcio de
  Investigaci\'on del Golfo de M\'exico (CIGoM). FAC thanks CICESE
  (Mexico) for allowing him to use their computer facilities
  throughout the CIGoM project.
\end{acknowledgments}

\section*{AIP Publishing data sharing policy}

The gridded multimission altimeter products were produced by
SSALTO/DUACS and distributed by \A\, (\texttt{https://\allowbreak
www.aviso.altimetry.fr/}), with support from CNES. The Mesoscale
Eddy Trajectory Atlas Product was produced by SSALTO/DUACS and
distributed by \A\, (\texttt{https://\allowbreak www.aviso.altimetry.fr/})
with support from CNES, in collaboration with Oregon State University
with support from NASA.  The \E\, data are available from \H.'s
website at \texttt{https://\allowbreak www.horizonmarine.com/}.


\begin{thebibliography}{38}%
\makeatletter
\providecommand \@ifxundefined [1]{%
 \@ifx{#1\undefined}
}%
\providecommand \@ifnum [1]{%
 \ifnum #1\expandafter \@firstoftwo
 \else \expandafter \@secondoftwo
 \fi
}%
\providecommand \@ifx [1]{%
 \ifx #1\expandafter \@firstoftwo
 \else \expandafter \@secondoftwo
 \fi
}%
\providecommand \natexlab [1]{#1}%
\providecommand \enquote  [1]{``#1''}%
\providecommand \bibnamefont  [1]{#1}%
\providecommand \bibfnamefont [1]{#1}%
\providecommand \citenamefont [1]{#1}%
\providecommand \href@noop [0]{\@secondoftwo}%
\providecommand \href [0]{\begingroup \@sanitize@url \@href}%
\providecommand \@href[1]{\@@startlink{#1}\@@href}%
\providecommand \@@href[1]{\endgroup#1\@@endlink}%
\providecommand \@sanitize@url [0]{\catcode `\\12\catcode `\$12\catcode
  `\&12\catcode `\#12\catcode `\^12\catcode `\_12\catcode `\%12\relax}%
\providecommand \@@startlink[1]{}%
\providecommand \@@endlink[0]{}%
\providecommand \url  [0]{\begingroup\@sanitize@url \@url }%
\providecommand \@url [1]{\endgroup\@href {#1}{\urlprefix }}%
\providecommand \urlprefix  [0]{URL }%
\providecommand \Eprint [0]{\href }%
\providecommand \doibase [0]{https://doi.org/}%
\providecommand \selectlanguage [0]{\@gobble}%
\providecommand \bibinfo  [0]{\@secondoftwo}%
\providecommand \bibfield  [0]{\@secondoftwo}%
\providecommand \translation [1]{[#1]}%
\providecommand \BibitemOpen [0]{}%
\providecommand \bibitemStop [0]{}%
\providecommand \bibitemNoStop [0]{.\EOS\space}%
\providecommand \EOS [0]{\spacefactor3000\relax}%
\providecommand \BibitemShut  [1]{\csname bibitem#1\endcsname}%
\let\auto@bib@innerbib\@empty
\bibitem [{\citenamefont {Andrade-Canto}, \citenamefont {Sheinbaum},\ and\
  \citenamefont {Sans\'on}(2013)}]{Andrade-etal-13}%
  \BibitemOpen
  \bibfield  {author} {\bibinfo {author} {\bibnamefont {Andrade-Canto},
  \bibfnamefont {F.}}, \bibinfo {author} {\bibnamefont {Sheinbaum},
  \bibfnamefont {J.}}, and\ \bibinfo {author} {\bibnamefont {Sans\'on},
  \bibfnamefont {L.~Z.}},\ }\bibfield  {title} {\enquote {\bibinfo {title} {{A
  Lagrangian approach to the Loop Current eddy separation}},}\ }\href@noop {}
  {\bibfield  {journal} {\bibinfo  {journal} {Nonlin. Processes Geophys.}\
  }\textbf {\bibinfo {volume} {20}},\ \bibinfo {pages} {85--96} (\bibinfo
  {year} {2013})}\BibitemShut {NoStop}%
\bibitem [{\citenamefont {Beron-Vera}\ \emph {et~al.}(2019)\citenamefont
  {Beron-Vera}, \citenamefont {Hadjighasem}, \citenamefont {Xia}, \citenamefont
  {Olascoaga},\ and\ \citenamefont {Haller}}]{Beron-etal-19-PNAS}%
  \BibitemOpen
  \bibfield  {author} {\bibinfo {author} {\bibnamefont {Beron-Vera},
  \bibfnamefont {F.~J.}}, \bibinfo {author} {\bibnamefont {Hadjighasem},
  \bibfnamefont {A.}}, \bibinfo {author} {\bibnamefont {Xia}, \bibfnamefont
  {Q.}}, \bibinfo {author} {\bibnamefont {Olascoaga}, \bibfnamefont {M.~J.}},
  and\ \bibinfo {author} {\bibnamefont {Haller}, \bibfnamefont {G.}},\
  }\bibfield  {title} {\enquote {\bibinfo {title} {{Coherent Lagrangian swirls
  among submesoscale motions}},}\ }\href@noop {} {\bibfield  {journal}
  {\bibinfo  {journal} {Proc. Natl. Acad. Sci. U.S.A.}\ }\textbf {\bibinfo
  {volume} {116}},\ \bibinfo {pages} {18251--18256} (\bibinfo {year}
  {2019})}\BibitemShut {NoStop}%
\bibitem [{\citenamefont {Beron-Vera}\ \emph {et~al.}(2018)\citenamefont
  {Beron-Vera}, \citenamefont {Olascaoaga}, \citenamefont {Wang}, \citenamefont
  {{Tri\~nanes}},\ and\ \citenamefont {{P\'erez}-Brunius}}]{Beron-etal-18}%
  \BibitemOpen
  \bibfield  {author} {\bibinfo {author} {\bibnamefont {Beron-Vera},
  \bibfnamefont {F.~J.}}, \bibinfo {author} {\bibnamefont {Olascaoaga},
  \bibfnamefont {M.~J.}}, \bibinfo {author} {\bibnamefont {Wang}, \bibfnamefont
  {Y.}}, \bibinfo {author} {\bibnamefont {{Tri\~nanes}}, \bibfnamefont {J.}},
  and\ \bibinfo {author} {\bibnamefont {{P\'erez}-Brunius}, \bibfnamefont
  {P.}},\ }\bibfield  {title} {\enquote {\bibinfo {title} {{Enduring Lagrangian
  coherence of a Loop Current ring assessed using independent observations}},}\
  }\href@noop {} {\bibfield  {journal} {\bibinfo  {journal} {Scientific
  Reports}\ }\textbf {\bibinfo {volume} {8}},\ \bibinfo {pages} {11275}
  (\bibinfo {year} {2018})}\BibitemShut {NoStop}%
\bibitem [{\citenamefont {Beron-Vera}\ \emph {et~al.}(2013)\citenamefont
  {Beron-Vera}, \citenamefont {Wang}, \citenamefont {Olascoaga}, \citenamefont
  {Goni},\ and\ \citenamefont {Haller}}]{Beron-etal-13}%
  \BibitemOpen
  \bibfield  {author} {\bibinfo {author} {\bibnamefont {Beron-Vera},
  \bibfnamefont {F.~J.}}, \bibinfo {author} {\bibnamefont {Wang}, \bibfnamefont
  {Y.}}, \bibinfo {author} {\bibnamefont {Olascoaga}, \bibfnamefont {M.~J.}},
  \bibinfo {author} {\bibnamefont {Goni}, \bibfnamefont {G.~J.}}, and\ \bibinfo
  {author} {\bibnamefont {Haller}, \bibfnamefont {G.}},\ }\bibfield  {title}
  {\enquote {\bibinfo {title} {{Objective detection of oceanic eddies and the
  Agulhas leakage}},}\ }\href@noop {} {\bibfield  {journal} {\bibinfo
  {journal} {J. Phys. Oceanogr.}\ }\textbf {\bibinfo {volume} {43}},\ \bibinfo
  {pages} {1426--1438} (\bibinfo {year} {2013})}\BibitemShut {NoStop}%
\bibitem [{\citenamefont {Chelton}, \citenamefont {Schlax},\ and\ \citenamefont
  {Samelson}(2011)}]{Chelton-etal-11a}%
  \BibitemOpen
  \bibfield  {author} {\bibinfo {author} {\bibnamefont {Chelton}, \bibfnamefont
  {D.~B.}}, \bibinfo {author} {\bibnamefont {Schlax}, \bibfnamefont {M.~G.}},
  and\ \bibinfo {author} {\bibnamefont {Samelson}, \bibfnamefont {R.~M.}},\
  }\bibfield  {title} {\enquote {\bibinfo {title} {Global observations of
  nonlinear mesoscale eddies},}\ }\href@noop {} {\bibfield  {journal} {\bibinfo
   {journal} {Prog. Oceanogr.}\ }\textbf {\bibinfo {volume} {91}},\ \bibinfo
  {pages} {167--216} (\bibinfo {year} {2011})}\BibitemShut {NoStop}%
\bibitem [{\citenamefont {Donohue}\ \emph {et~al.}(2016)\citenamefont
  {Donohue}, \citenamefont {Watts}, \citenamefont {Hamilton}, \citenamefont
  {Leben},\ and\ \citenamefont {Kennelly}}]{Donohue-etal-16}%
  \BibitemOpen
  \bibfield  {author} {\bibinfo {author} {\bibnamefont {Donohue}, \bibfnamefont
  {K.}}, \bibinfo {author} {\bibnamefont {Watts}, \bibfnamefont {D.}}, \bibinfo
  {author} {\bibnamefont {Hamilton}, \bibfnamefont {P.}}, \bibinfo {author}
  {\bibnamefont {Leben}, \bibfnamefont {R.}}, and\ \bibinfo {author}
  {\bibnamefont {Kennelly}, \bibfnamefont {M.}},\ }\bibfield  {title} {\enquote
  {\bibinfo {title} {Loop current eddy formation and baroclinic instability},}\
  }\href {https://doi.org/https://doi.org/10.1016/j.dynatmoce.2016.01.004}
  {\bibfield  {journal} {\bibinfo  {journal} {Dynamics of Atmospheres and
  Oceans}\ }\textbf {\bibinfo {volume} {76}},\ \bibinfo {pages} {195--216}
  (\bibinfo {year} {2016})},\ \bibinfo {note} {the Loop Current Dynamics
  Experiment}\BibitemShut {NoStop}%
\bibitem [{\citenamefont {Forristall}, \citenamefont {Schaudt},\ and\
  \citenamefont {Cooper}(1992)}]{Forristall-etal-92}%
  \BibitemOpen
  \bibfield  {author} {\bibinfo {author} {\bibnamefont {Forristall},
  \bibfnamefont {G.~Z.}}, \bibinfo {author} {\bibnamefont {Schaudt},
  \bibfnamefont {K.~J.}}, and\ \bibinfo {author} {\bibnamefont {Cooper},
  \bibfnamefont {C.~K.}},\ }\bibfield  {title} {\enquote {\bibinfo {title}
  {Evolution and kinematics of a {L}oop {C}urrent eddy in the {G}ulf of
  {M}exico during 1985},}\ }\href@noop {} {\bibfield  {journal} {\bibinfo
  {journal} {J. Geophys. Res.}\ }\textbf {\bibinfo {volume} {97}},\ \bibinfo
  {pages} {2173--2184} (\bibinfo {year} {1992})}\BibitemShut {NoStop}%
\bibitem [{\citenamefont {Hadjighasem}\ \emph {et~al.}(2017)\citenamefont
  {Hadjighasem}, \citenamefont {Farazmand}, \citenamefont {Blazevski},
  \citenamefont {Froyland},\ and\ \citenamefont
  {Haller}}]{Hadjighasem-etal-17}%
  \BibitemOpen
  \bibfield  {author} {\bibinfo {author} {\bibnamefont {Hadjighasem},
  \bibfnamefont {A.}}, \bibinfo {author} {\bibnamefont {Farazmand},
  \bibfnamefont {M.}}, \bibinfo {author} {\bibnamefont {Blazevski},
  \bibfnamefont {D.}}, \bibinfo {author} {\bibnamefont {Froyland},
  \bibfnamefont {G.}}, and\ \bibinfo {author} {\bibnamefont {Haller},
  \bibfnamefont {G.}},\ }\bibfield  {title} {\enquote {\bibinfo {title} {A
  critical comparison of {L}agrangian methods for coherent structure
  detection},}\ }\href@noop {} {\bibfield  {journal} {\bibinfo  {journal}
  {Chaos}\ }\textbf {\bibinfo {volume} {27}},\ \bibinfo {pages} {053104}
  (\bibinfo {year} {2017})}\BibitemShut {NoStop}%
\bibitem [{\citenamefont {Haller}(2005)}]{Haller-05}%
  \BibitemOpen
  \bibfield  {author} {\bibinfo {author} {\bibnamefont {Haller}, \bibfnamefont
  {G.}},\ }\bibfield  {title} {\enquote {\bibinfo {title} {An objective
  definition of a vortex},}\ }\href {https://doi.org/10.1017/S0022112004002526}
  {\bibfield  {journal} {\bibinfo  {journal} {J. Fluid Mech.}\ }\textbf
  {\bibinfo {volume} {525}},\ \bibinfo {pages} {1--26} (\bibinfo {year}
  {2005})}\BibitemShut {NoStop}%
\bibitem [{\citenamefont {Haller}(2015)}]{Haller-15}%
  \BibitemOpen
  \bibfield  {author} {\bibinfo {author} {\bibnamefont {Haller}, \bibfnamefont
  {G.}},\ }\bibfield  {title} {\enquote {\bibinfo {title} {Lagrangian coherent
  structures},}\ }\href {https://doi.org/10.1146/annurev-fluid-010313-141322}
  {\bibfield  {journal} {\bibinfo  {journal} {Ann. Rev. Fluid Mech.}\ }\textbf
  {\bibinfo {volume} {47}},\ \bibinfo {pages} {137--162} (\bibinfo {year}
  {2015})}\BibitemShut {NoStop}%
\bibitem [{\citenamefont {Haller}(2016)}]{Haller-16}%
  \BibitemOpen
  \bibfield  {author} {\bibinfo {author} {\bibnamefont {Haller}, \bibfnamefont
  {G.}},\ }\bibfield  {title} {\enquote {\bibinfo {title} {Climate, black holes
  and vorticity: {H}ow on {E}arth are they related?}}\ }\href@noop {}
  {\bibfield  {journal} {\bibinfo  {journal} {SIAM News}\ }\textbf {\bibinfo
  {volume} {49}},\ \bibinfo {pages} {1--2} (\bibinfo {year}
  {2016})}\BibitemShut {NoStop}%
\bibitem [{\citenamefont {Haller}\ and\ \citenamefont
  {Beron-Vera}(2012)}]{Haller-Beron-12}%
  \BibitemOpen
  \bibfield  {author} {\bibinfo {author} {\bibnamefont {Haller}, \bibfnamefont
  {G.}}and\ \bibinfo {author} {\bibnamefont {Beron-Vera}, \bibfnamefont
  {F.~J.}},\ }\bibfield  {title} {\enquote {\bibinfo {title} {Geodesic theory
  of transport barriers in two-dimensional flows},}\ }\href
  {https://doi.org/10.1016/j.physd.2012.06.012} {\bibfield  {journal} {\bibinfo
   {journal} {Physica D}\ }\textbf {\bibinfo {volume} {241}},\ \bibinfo {pages}
  {1680--1702} (\bibinfo {year} {2012})}\BibitemShut {NoStop}%
\bibitem [{\citenamefont {Haller}\ and\ \citenamefont
  {Beron-Vera}(2013)}]{Haller-Beron-13}%
  \BibitemOpen
  \bibfield  {author} {\bibinfo {author} {\bibnamefont {Haller}, \bibfnamefont
  {G.}}and\ \bibinfo {author} {\bibnamefont {Beron-Vera}, \bibfnamefont
  {F.~J.}},\ }\bibfield  {title} {\enquote {\bibinfo {title} {{Coherent
  Lagrangian vortices: The black holes of turbulence}},}\ }\href@noop {}
  {\bibfield  {journal} {\bibinfo  {journal} {J. Fluid Mech.}\ }\textbf
  {\bibinfo {volume} {731}},\ \bibinfo {pages} {R4} (\bibinfo {year}
  {2013})}\BibitemShut {NoStop}%
\bibitem [{\citenamefont {Haller}\ and\ \citenamefont
  {Beron-Vera}(2014)}]{Haller-Beron-14}%
  \BibitemOpen
  \bibfield  {author} {\bibinfo {author} {\bibnamefont {Haller}, \bibfnamefont
  {G.}}and\ \bibinfo {author} {\bibnamefont {Beron-Vera}, \bibfnamefont
  {F.~J.}},\ }\bibfield  {title} {\enquote {\bibinfo {title} {{Addendum to
  `Coherent Lagrangian vortices: The black holes of turbulence'}},}\
  }\href@noop {} {\bibfield  {journal} {\bibinfo  {journal} {J. Fluid Mech.}\
  }\textbf {\bibinfo {volume} {755}},\ \bibinfo {pages} {R3} (\bibinfo {year}
  {2014})}\BibitemShut {NoStop}%
\bibitem [{\citenamefont {Haller}\ \emph {et~al.}(2016)\citenamefont {Haller},
  \citenamefont {Hadjighasem}, \citenamefont {Farazmand},\ and\ \citenamefont
  {Huhn}}]{Haller-etal-16}%
  \BibitemOpen
  \bibfield  {author} {\bibinfo {author} {\bibnamefont {Haller}, \bibfnamefont
  {G.}}, \bibinfo {author} {\bibnamefont {Hadjighasem}, \bibfnamefont {A.}},
  \bibinfo {author} {\bibnamefont {Farazmand}, \bibfnamefont {M.}}, and\
  \bibinfo {author} {\bibnamefont {Huhn}, \bibfnamefont {F.}},\ }\bibfield
  {title} {\enquote {\bibinfo {title} {Defining coherent vortices objectively
  from the vorticity},}\ }\href@noop {} {\bibfield  {journal} {\bibinfo
  {journal} {J. Fluid Mech.}\ }\textbf {\bibinfo {volume} {795}},\ \bibinfo
  {pages} {136--173} (\bibinfo {year} {2016})}\BibitemShut {NoStop}%
\bibitem [{\citenamefont {Haller}, \citenamefont {Karrasch},\ and\
  \citenamefont {Kogelbauer}(2018)}]{Haller-etal-18}%
  \BibitemOpen
  \bibfield  {author} {\bibinfo {author} {\bibnamefont {Haller}, \bibfnamefont
  {G.}}, \bibinfo {author} {\bibnamefont {Karrasch}, \bibfnamefont {D.}}, and\
  \bibinfo {author} {\bibnamefont {Kogelbauer}, \bibfnamefont {F.}},\
  }\bibfield  {title} {\enquote {\bibinfo {title} {Material barriers to
  diffusive and stochastic transport},}\ }\href@noop {} {\bibfield  {journal}
  {\bibinfo  {journal} {Proceedings of the National Academy of Sciences}\
  }\textbf {\bibinfo {volume} {115}},\ \bibinfo {pages} {9074--9079} (\bibinfo
  {year} {2018})}\BibitemShut {NoStop}%
\bibitem [{\citenamefont {Haller}, \citenamefont {Karrasch},\ and\
  \citenamefont {Kogelbauer}(2020)}]{Haller-etal-19}%
  \BibitemOpen
  \bibfield  {author} {\bibinfo {author} {\bibnamefont {Haller}, \bibfnamefont
  {G.}}, \bibinfo {author} {\bibnamefont {Karrasch}, \bibfnamefont {D.}}, and\
  \bibinfo {author} {\bibnamefont {Kogelbauer}, \bibfnamefont {F.}},\
  }\bibfield  {title} {\enquote {\bibinfo {title} {Barriers to the transport of
  diffusive scalars in compressible flows},}\ }\href@noop {} {\bibfield
  {journal} {\bibinfo  {journal} {SIAM Journal on Applied Dynamical Systems}\
  }\textbf {\bibinfo {volume} {19}},\ \bibinfo {pages} {85--123} (\bibinfo
  {year} {2020})}\BibitemShut {NoStop}%
\bibitem [{\citenamefont {Kantha}(2014)}]{Kantha-14}%
  \BibitemOpen
  \bibfield  {author} {\bibinfo {author} {\bibnamefont {Kantha}, \bibfnamefont
  {L.}},\ }\bibfield  {title} {\enquote {\bibinfo {title} {Empirical models of
  the loop current eddy detachment/separation time in the gulf of mexico},}\
  }\href@noop {} {\bibfield  {journal} {\bibinfo  {journal} {Journal of
  Waterway, Port, Coastal, and Ocean Engineering}\ }\textbf {\bibinfo {volume}
  {140}},\ \bibinfo {pages} {04014001} (\bibinfo {year} {2014})}\BibitemShut
  {NoStop}%
\bibitem [{\citenamefont {Karrasch}, \citenamefont {Huhn},\ and\ \citenamefont
  {Haller}(2014)}]{Karrasch-etal-14}%
  \BibitemOpen
  \bibfield  {author} {\bibinfo {author} {\bibnamefont {Karrasch},
  \bibfnamefont {D.}}, \bibinfo {author} {\bibnamefont {Huhn}, \bibfnamefont
  {F.}}, and\ \bibinfo {author} {\bibnamefont {Haller}, \bibfnamefont {G.}},\
  }\bibfield  {title} {\enquote {\bibinfo {title} {{Automated detection of
  coherent Lagrangian vortices in two-dimensional unsteady flows}},}\
  }\href@noop {} {\bibfield  {journal} {\bibinfo  {journal} {Proc. Royal Soc.
  A}\ }\textbf {\bibinfo {volume} {471}},\ \bibinfo {pages} {20140639}
  (\bibinfo {year} {2014})}\BibitemShut {NoStop}%
\bibitem [{\citenamefont {Karrasch}\ and\ \citenamefont
  {Schilling}(2020)}]{Karrasch-Schilling-20}%
  \BibitemOpen
  \bibfield  {author} {\bibinfo {author} {\bibnamefont {Karrasch},
  \bibfnamefont {D.}}and\ \bibinfo {author} {\bibnamefont {Schilling},
  \bibfnamefont {N.}},\ }\bibfield  {title} {\enquote {\bibinfo {title} {Fast
  and robust computation of coherent lagrangian vortices on very large
  two-dimensional domains},}\ }\href@noop {} {\bibfield  {journal} {\bibinfo
  {journal} {The SMAI journal of computational mathematics}\ }\textbf {\bibinfo
  {volume} {6}},\ \bibinfo {pages} {101--124} (\bibinfo {year}
  {2020})}\BibitemShut {NoStop}%
\bibitem [{\citenamefont {Kuznetsov}\ \emph {et~al.}(2002)\citenamefont
  {Kuznetsov}, \citenamefont {Toner}, \citenamefont {Kirwan}, \citenamefont
  {Jones}, \citenamefont {Kantha},\ and\ \citenamefont
  {Choi}}]{Kuznetsov-etal-02}%
  \BibitemOpen
  \bibfield  {author} {\bibinfo {author} {\bibnamefont {Kuznetsov},
  \bibfnamefont {L.}}, \bibinfo {author} {\bibnamefont {Toner}, \bibfnamefont
  {M.}}, \bibinfo {author} {\bibnamefont {Kirwan}, \bibfnamefont {A.~D.}},
  \bibinfo {author} {\bibnamefont {Jones}, \bibfnamefont {C.~K. R.~T.}},
  \bibinfo {author} {\bibnamefont {Kantha}, \bibfnamefont {L.~H.}}, and\
  \bibinfo {author} {\bibnamefont {Choi}, \bibfnamefont {J.}},\ }\bibfield
  {title} {\enquote {\bibinfo {title} {The {L}oop {C}urrent and adjacent rings
  delineated by {L}agrangian analysis of the near-surface flow},}\ }\href@noop
  {} {\bibfield  {journal} {\bibinfo  {journal} {J. Mar. Res.}\ }\textbf
  {\bibinfo {volume} {60}},\ \bibinfo {pages} {405--429} (\bibinfo {year}
  {2002})}\BibitemShut {NoStop}%
\bibitem [{\citenamefont {{Le Traon}}, \citenamefont {Nadal},\ and\
  \citenamefont {Ducet}(1998)}]{LeTraon-etal-98}%
  \BibitemOpen
  \bibfield  {author} {\bibinfo {author} {\bibnamefont {{Le Traon}},
  \bibfnamefont {P.~Y.}}, \bibinfo {author} {\bibnamefont {Nadal},
  \bibfnamefont {F.}}, and\ \bibinfo {author} {\bibnamefont {Ducet},
  \bibfnamefont {N.}},\ }\bibfield  {title} {\enquote {\bibinfo {title} {An
  improved mapping method of multisatellite altimeter data},}\ }\href@noop {}
  {\bibfield  {journal} {\bibinfo  {journal} {J. Atmos. Oceanic Technol.}\
  }\textbf {\bibinfo {volume} {15}},\ \bibinfo {pages} {522--534} (\bibinfo
  {year} {1998})}\BibitemShut {NoStop}%
\bibitem [{\citenamefont {Leben}(2005)}]{Leben-05}%
  \BibitemOpen
  \bibfield  {author} {\bibinfo {author} {\bibnamefont {Leben}, \bibfnamefont
  {R.~R.}},\ }\bibfield  {title} {\enquote {\bibinfo {title} {Altimeter-derived
  loop current metrics},}\ }in\ \href {https://doi.org/10.1029/161GM15} {\emph
  {\bibinfo {booktitle} {Circulation in the Gulf of Mexico: Observations and
  Models}}},\ \bibinfo {editor} {edited by\ \bibinfo {editor} {\bibfnamefont
  {W.}~\bibnamefont {Sturges}}\ and\ \bibinfo {editor} {\bibfnamefont
  {A.}~\bibnamefont {Lugo-Fernandez}}}\ (\bibinfo  {publisher} {American
  Geophysical Union},\ \bibinfo {year} {2005})\ pp.\ \bibinfo {pages}
  {181--201}\BibitemShut {NoStop}%
\bibitem [{\citenamefont {Lewis}\ and\ \citenamefont
  {Kirwan}(1987)}]{Lewis-Kirwan-87}%
  \BibitemOpen
  \bibfield  {author} {\bibinfo {author} {\bibnamefont {Lewis}, \bibfnamefont
  {J.~K.}}and\ \bibinfo {author} {\bibnamefont {Kirwan}, \bibfnamefont
  {A.~D.}},\ }\bibfield  {title} {\enquote {\bibinfo {title} {{Genesis of a
  Gulf of Mexico ring as determined from kinematic analyses}},}\ }\href@noop {}
  {\bibfield  {journal} {\bibinfo  {journal} {J. Geophys. Res.}\ }\textbf
  {\bibinfo {volume} {92}},\ \bibinfo {pages} {11727--11740} (\bibinfo {year}
  {1987})}\BibitemShut {NoStop}%
\bibitem [{\citenamefont {Lipphardt}\ \emph {et~al.}(2008)\citenamefont
  {Lipphardt}, \citenamefont {Poje}, \citenamefont {Kirwan}, \citenamefont
  {Kantha},\ and\ \citenamefont {Zweng}}]{Lipphardt-etal-08}%
  \BibitemOpen
  \bibfield  {author} {\bibinfo {author} {\bibnamefont {Lipphardt},
  \bibfnamefont {B.~L.}}, \bibinfo {author} {\bibnamefont {Poje}, \bibfnamefont
  {A.~C.}}, \bibinfo {author} {\bibnamefont {Kirwan}, \bibfnamefont {A.~D.}},
  \bibinfo {author} {\bibnamefont {Kantha}, \bibfnamefont {L.}}, and\ \bibinfo
  {author} {\bibnamefont {Zweng}, \bibfnamefont {W.}},\ }\bibfield  {title}
  {\enquote {\bibinfo {title} {{Death of three Loop Current rings}},}\
  }\href@noop {} {\bibfield  {journal} {\bibinfo  {journal} {J. Marine Res.}\
  }\textbf {\bibinfo {volume} {66}},\ \bibinfo {pages} {25--60} (\bibinfo
  {year} {2008})}\BibitemShut {NoStop}%
\bibitem [{\citenamefont {Lugt}(1979)}]{Lugt-79}%
  \BibitemOpen
  \bibfield  {author} {\bibinfo {author} {\bibnamefont {Lugt}, \bibfnamefont
  {H.~J.}},\ }\bibfield  {title} {\enquote {\bibinfo {title} {The dilemma of
  defining a vortex},}\ }in\ \href@noop {} {\emph {\bibinfo {booktitle} {Recent
  Developments in Theoretical and Experimental Fluid Mechanics}}},\ \bibinfo
  {editor} {edited by\ \bibinfo {editor} {\bibfnamefont {U.}~\bibnamefont
  {Muller}}, \bibinfo {editor} {\bibfnamefont {K.~G.}\ \bibnamefont {Riesner}},
  \ and\ \bibinfo {editor} {\bibfnamefont {B.}~\bibnamefont {Schmidt}}}\
  (\bibinfo  {publisher} {Springer-Verlag},\ \bibinfo {year} {1979})\ pp.\
  \bibinfo {pages} {309--321}\BibitemShut {NoStop}%
\bibitem [{\citenamefont {Okubo}(1970)}]{Okubo-70}%
  \BibitemOpen
  \bibfield  {author} {\bibinfo {author} {\bibnamefont {Okubo}, \bibfnamefont
  {A.}},\ }\bibfield  {title} {\enquote {\bibinfo {title} {Horizontal
  dispersion of flotable particles in the vicinity of velocity singularity such
  as convergences},}\ }\href@noop {} {\bibfield  {journal} {\bibinfo  {journal}
  {Deep-Sea Res. Oceanogr. Abstr.}\ }\textbf {\bibinfo {volume} {12}},\
  \bibinfo {pages} {445--454} (\bibinfo {year} {1970})}\BibitemShut {NoStop}%
\bibitem [{\citenamefont {Olascoaga}\ \emph {et~al.}(2013)\citenamefont
  {Olascoaga}, \citenamefont {Beron-Vera}, \citenamefont {Haller},
  \citenamefont {Trinanes}, \citenamefont {Iskandarani}, \citenamefont
  {Coelho}, \citenamefont {Haus}, \citenamefont {H.~S.~Huntley}, \citenamefont
  {D.{Kirwan, Jr.}}, \citenamefont {{Lipphardt, Jr.}}, \citenamefont
  {\"{O}zg\"{o}kmen}, \citenamefont {Reniers},\ and\ \citenamefont
  {Valle-Levinson}}]{Olascoaga-etal-13}%
  \BibitemOpen
  \bibfield  {author} {\bibinfo {author} {\bibnamefont {Olascoaga},
  \bibfnamefont {M.~J.}}, \bibinfo {author} {\bibnamefont {Beron-Vera},
  \bibfnamefont {F.~J.}}, \bibinfo {author} {\bibnamefont {Haller},
  \bibfnamefont {G.}}, \bibinfo {author} {\bibnamefont {Trinanes},
  \bibfnamefont {J.}}, \bibinfo {author} {\bibnamefont {Iskandarani},
  \bibfnamefont {M.}}, \bibinfo {author} {\bibnamefont {Coelho}, \bibfnamefont
  {E.~F.}}, \bibinfo {author} {\bibnamefont {Haus}, \bibfnamefont {B.}},
  \bibinfo {author} {\bibnamefont {H.~S.~Huntley}, \bibfnamefont {G.~J.}},
  \bibinfo {author} {\bibnamefont {D.{Kirwan, Jr.}}, \bibfnamefont {A.}},
  \bibinfo {author} {\bibnamefont {{Lipphardt, Jr.}}, \bibfnamefont {B.~L.}},
  \bibinfo {author} {\bibnamefont {\"{O}zg\"{o}kmen}, \bibfnamefont {T.}},
  \bibinfo {author} {\bibnamefont {Reniers}, \bibfnamefont {A.~J.~H.~M.}}, and\
  \bibinfo {author} {\bibnamefont {Valle-Levinson}, \bibfnamefont {A.}},\
  }\bibfield  {title} {\enquote {\bibinfo {title} {{Drifter motion in the Gulf
  of Mexico constrained by altimetric Lagrangian Coherent Structures}},}\
  }\href@noop {} {\bibfield  {journal} {\bibinfo  {journal} {Geophys. Res.
  Lett.}\ }\textbf {\bibinfo {volume} {40}},\ \bibinfo {pages} {6171--6175}
  (\bibinfo {year} {2013})}\BibitemShut {NoStop}%
\bibitem [{\citenamefont {Peacock}, \citenamefont {Froyland},\ and\
  \citenamefont {Haller}(2015)}]{Peacock-etal-15}%
  \BibitemOpen
  \bibfield  {author} {\bibinfo {author} {\bibnamefont {Peacock}, \bibfnamefont
  {T.}}, \bibinfo {author} {\bibnamefont {Froyland}, \bibfnamefont {G.}}, and\
  \bibinfo {author} {\bibnamefont {Haller}, \bibfnamefont {G.}},\ }\bibfield
  {title} {\enquote {\bibinfo {title} {Introduction to focus issue: Objective
  detection of coherent structures},}\ }\href@noop {} {\bibfield  {journal}
  {\bibinfo  {journal} {Chaos}\ }\textbf {\bibinfo {volume} {25}},\ \bibinfo
  {pages} {087201} (\bibinfo {year} {2015})}\BibitemShut {NoStop}%
\bibitem [{\citenamefont {Rio}\ and\ \citenamefont
  {Hernandez}(2004)}]{Rio-Hernandez-04}%
  \BibitemOpen
  \bibfield  {author} {\bibinfo {author} {\bibnamefont {Rio}, \bibfnamefont
  {M.-H.}}and\ \bibinfo {author} {\bibnamefont {Hernandez}, \bibfnamefont
  {F.}},\ }\bibfield  {title} {\enquote {\bibinfo {title} {A mean dynamic
  topography computed over the world ocean from altimetry, in situ measurements
  and a geoid model},}\ }\href {https://doi.org/10.1029/2003JC002226}
  {\bibfield  {journal} {\bibinfo  {journal} {J. Geophys. Res.}\ }\textbf
  {\bibinfo {volume} {109}},\ \bibinfo {pages} {C12032} (\bibinfo {year}
  {2004})}\BibitemShut {NoStop}%
\bibitem [{\citenamefont {Serra}\ and\ \citenamefont
  {Haller}(2016)}]{Serra-Haller-16}%
  \BibitemOpen
  \bibfield  {author} {\bibinfo {author} {\bibnamefont {Serra}, \bibfnamefont
  {M.}}and\ \bibinfo {author} {\bibnamefont {Haller}, \bibfnamefont {G.}},\
  }\bibfield  {title} {\enquote {\bibinfo {title} {{Objective Eulerian coherent
  structures}},}\ }\href@noop {} {\bibfield  {journal} {\bibinfo  {journal}
  {Chaos}\ }\textbf {\bibinfo {volume} {26}},\ \bibinfo {pages} {053110}
  (\bibinfo {year} {2016})}\BibitemShut {NoStop}%
\bibitem [{\citenamefont {Shay}, \citenamefont {Goni},\ and\ \citenamefont
  {Black}(2000)}]{Shay-etal-00}%
  \BibitemOpen
  \bibfield  {author} {\bibinfo {author} {\bibnamefont {Shay}, \bibfnamefont
  {L.~K.}}, \bibinfo {author} {\bibnamefont {Goni}, \bibfnamefont {G.~J.}},
  and\ \bibinfo {author} {\bibnamefont {Black}, \bibfnamefont {P.~G.}},\
  }\bibfield  {title} {\enquote {\bibinfo {title} {Effects of a warm oceanic
  feature on hurricane opal},}\ }\href@noop {} {\bibfield  {journal} {\bibinfo
  {journal} {Mon. Weather Rev.}\ }\textbf {\bibinfo {volume} {128}},\ \bibinfo
  {pages} {1366--1383} (\bibinfo {year} {2000})}\BibitemShut {NoStop}%
\bibitem [{\citenamefont {Sturges}\ and\ \citenamefont
  {Leben}(2000)}]{Sturges-Leben-00}%
  \BibitemOpen
  \bibfield  {author} {\bibinfo {author} {\bibnamefont {Sturges}, \bibfnamefont
  {W.}}and\ \bibinfo {author} {\bibnamefont {Leben}, \bibfnamefont {R.}},\
  }\bibfield  {title} {\enquote {\bibinfo {title} {{Frequency of ring
  separation from the Lopp Current in the Gulf of Mexico: A revised
  estimate}},}\ }\href@noop {} {\bibfield  {journal} {\bibinfo  {journal} {J.
  Phys. Oceanogr.}\ }\textbf {\bibinfo {volume} {30}},\ \bibinfo {pages}
  {1814--1819} (\bibinfo {year} {2000})}\BibitemShut {NoStop}%
\bibitem [{\citenamefont {Sturges}\ and\ \citenamefont
  {Lugo-Fernandez}(2005)}]{Sturges-05}%
  \BibitemOpen
  \bibfield  {author} {\bibinfo {author} {\bibnamefont {Sturges}, \bibfnamefont
  {W.}}and\ \bibinfo {author} {\bibnamefont {Lugo-Fernandez}, \bibfnamefont
  {A.}},\ }\bibfield  {title} {\enquote {\bibinfo {title} {Circulation in the
  gulf of mexico: Observations and models},}\ }\href@noop {} {\bibfield
  {journal} {\bibinfo  {journal} {Washington DC American Geophysical Union
  Geophysical Monograph Series}\ }\textbf {\bibinfo {volume} {161}} (\bibinfo
  {year} {2005})}\BibitemShut {NoStop}%
\bibitem [{\citenamefont {Tarshish}\ \emph {et~al.}(2018)\citenamefont
  {Tarshish}, \citenamefont {Abernathey}, \citenamefont {Zhang}, \citenamefont
  {Dufour}, \citenamefont {Frenger},\ and\ \citenamefont
  {Griffies}}]{Tarshish-etal-18}%
  \BibitemOpen
  \bibfield  {author} {\bibinfo {author} {\bibnamefont {Tarshish},
  \bibfnamefont {N.}}, \bibinfo {author} {\bibnamefont {Abernathey},
  \bibfnamefont {R.}}, \bibinfo {author} {\bibnamefont {Zhang}, \bibfnamefont
  {C.}}, \bibinfo {author} {\bibnamefont {Dufour}, \bibfnamefont {C.~O.}},
  \bibinfo {author} {\bibnamefont {Frenger}, \bibfnamefont {I.}}, and\ \bibinfo
  {author} {\bibnamefont {Griffies}, \bibfnamefont {S.~M.}},\ }\bibfield
  {title} {\enquote {\bibinfo {title} {Identifying lagrangian coherent vortices
  in a mesoscale ocean model},}\ }\href@noop {} {\bibfield  {journal} {\bibinfo
   {journal} {Ocean Modelling}\ }\textbf {\bibinfo {volume} {130}},\ \bibinfo
  {pages} {15 -- 28} (\bibinfo {year} {2018})}\BibitemShut {NoStop}%
\bibitem [{\citenamefont {Wang}, \citenamefont {Olascoaga},\ and\ \citenamefont
  {Beron-Vera}(2015)}]{Wang-etal-15}%
  \BibitemOpen
  \bibfield  {author} {\bibinfo {author} {\bibnamefont {Wang}, \bibfnamefont
  {Y.}}, \bibinfo {author} {\bibnamefont {Olascoaga}, \bibfnamefont {M.~J.}},
  and\ \bibinfo {author} {\bibnamefont {Beron-Vera}, \bibfnamefont {F.~J.}},\
  }\bibfield  {title} {\enquote {\bibinfo {title} {{Coherent water transport
  across the South Atlantic}},}\ }\href {https://doi.org/10.1002/2015GL064089}
  {\bibfield  {journal} {\bibinfo  {journal} {Geophys. Res. Lett.}\ }\textbf
  {\bibinfo {volume} {42}},\ \bibinfo {pages} {4072--4079} (\bibinfo {year}
  {2015})}\BibitemShut {NoStop}%
\bibitem [{\citenamefont {Wang}, \citenamefont {Olascoaga},\ and\ \citenamefont
  {Beron-Vera}(2016)}]{Wang-etal-16}%
  \BibitemOpen
  \bibfield  {author} {\bibinfo {author} {\bibnamefont {Wang}, \bibfnamefont
  {Y.}}, \bibinfo {author} {\bibnamefont {Olascoaga}, \bibfnamefont {M.~J.}},
  and\ \bibinfo {author} {\bibnamefont {Beron-Vera}, \bibfnamefont {F.~J.}},\
  }\bibfield  {title} {\enquote {\bibinfo {title} {{The life cycle of a
  coherent Lagrangian Agulhas ring}},}\ }\href
  {https://doi.org/10.1002/2015JC011620} {\bibfield  {journal} {\bibinfo
  {journal} {J. Geophys. Res.}\ }\textbf {\bibinfo {volume} {121}},\ \bibinfo
  {pages} {3944--3954} (\bibinfo {year} {2016})}\BibitemShut {NoStop}%
\bibitem [{\citenamefont {Weiss}(1991)}]{Weiss-91}%
  \BibitemOpen
  \bibfield  {author} {\bibinfo {author} {\bibnamefont {Weiss}, \bibfnamefont
  {J.}},\ }\bibfield  {title} {\enquote {\bibinfo {title} {{The dynamics of
  enstrophy transfer in two-dimensional hydrodynamics}},}\ }\href
  {https://doi.org/10.1016/0167-2789(91)90088-Q} {\bibfield  {journal}
  {\bibinfo  {journal} {Physica D}\ }\textbf {\bibinfo {volume} {48}},\
  \bibinfo {pages} {273--294} (\bibinfo {year} {1991})}\BibitemShut {NoStop}%
\end{thebibliography}
%

\end{document}